# Title: Anticorrelated stereodynamics of scattering and sticking of $H_2$ molecules colliding with a reactive surface.

**Authors:** H. Chadwick[1], G. Zhang[1], C. J. Baker[1], P. L. Smith[1] and G. Alexandrowicz[1*]

**Affiliations:**

[1]Department of Chemistry, Faculty of Science and Engineering, Swansea University, Swansea SA2 8PP, UK.

*Corresponding author: g.n.alexandrowicz@swansea.ac.uk

**Abstract:** *When hydrogen molecules collide with a surface, they can either scatter away from the surface or stick to the surface through a dissociation reaction which leaves two H atoms adsorbed on the surface. The relative probabilities of these two potential outcomes can depend on the rotational orientation of the impinging molecules, however, direct measurements of this dependence were not available due to the difficulty of controlling the rotational orientation of ground state $H_2$ molecules. Here, we use magnetic manipulation to achieve rotational orientation control of the molecules just before they collide with the surface, and show that molecules approaching the surface in a helicopter orientation have a higher probability to react and dissociate, whereas those which approach in a cartwheel orientation are more likely to scatter.*

**Main Text:** Controlling the quantum state of reagents in a chemical reaction and observing stereodynamic (orientation-dependent) effects in the reaction rate, provides an ultimate test for our understanding of chemistry at the single molecule level(*1*). Various methods have been successfully applied over the years to achieve stereodynamic control, using photo-excitation, trajectory deflection in electric and magnetic field gradients and velocity-rotational orientation correlations in nozzle expansions (e.g.(*2–8*)).

The stereodynamic control methods mentioned above work for specific molecules in specific quantum states. One molecule they cannot address is $H_2$ when it is in the vibrational and rotational ground state. Beyond being the most abundant molecule in the universe, $H_2$ is the smallest and simplest molecular reagent, allowing relatively accurate theoretical modelling of gas phase and molecule-surface collisions, and also plays a role in numerous processes ranging from astrochemistry to the development of clean renewable energy(*9–12*). In this study we used a modified Magnetic Molecular Interferometer (MMI) setup(*13, 14*) to control the rotational projection quantum state of ground state $H_2$ molecules, and modulate the rotational orientation of the molecules just before they collide with a Ni(111) surface. We monitored two collisional processes; specular scattering back into the gas phase and dissociative chemisorption. Clear stereodynamic trends were seen in both processes.

The dissociative adsorption reaction of $H_2$ on Ni(111) has been shown by several molecular beam experiments to be an activated process, with sticking coefficients which increase linearly with the energy of the impinging molecules(*15, 16*), trends which can be qualitatively reproduced by semi-empirical density function theory(*17*). There are reasons to believe stereodynamic effects might play a role in the interaction. For example, $D_2$ molecules were observed to desorb from Cu(111) preferentially with their rotation plane parallel to the surface, indicating a similar preference for the reverse reaction(*18*), and calculations have predicted stereodynamic trends of $H_2$ dissociation on both flat and stepped copper surfaces(*19, 20*). Furthermore, theoretical work has predicted a direct relation between the stereodynamics of scattering and dissociation for $H_2$ colliding with a cobalt surface(*21*), however, it has previously not been possible to probe such correlations experimentally.

In order to study the stereodynamic trends of $H_2$ molecules colliding with a Ni(111) surface, we modified the MMI setup(*13, 14*), adding two differentially-pumped mass spectrometers which allow us to measure the total scattered flux at two angles, in a way which is independent of the final quantum state of the molecule. We denote this mode of operation as flux detection (FD) measurements. This is different and simpler than the configuration used in previous MMI experiments, where an additional hexapole was used before the detector, leading to a signal which depends on the nuclear spin and rotational projection states of the scattered molecules(*13, 14, 22*). The simplified version of the MMI setup we used for FD measurements, is illustrated schematically in figure 1a and described in more detail in the supplementary text.

The basic principle of our control scheme exploits the fact a ground state ortho-hydrogen molecule has 9 different eigen-energies which differ in their nuclear spin and rotational projections ($m_I, m_J$)(*23*). While the magnetic moments associated with the $m_I, m_J$ states of $H_2$ are 3 orders of magnitude weaker than those in paramagnetic, radical and metastable species(*4, 24, 25*), the subtle energy differences can still be used for state selective manipulation. More specifically, a hexapole polariser magnet(*26*) is used as a magnetic lens to create an initial population difference of quantum states in the beam. The hexapole magnet primarily selects the states based on their $m_I$ projections due to the large difference between the rotational and nuclear spin magnetic moments. The molecules then enter a perpendicular homogenous magnetic field, $B_1$, created by passing a current, $I_1$, through a solenoid coil, before entering the scattering chamber and colliding with the surface. The non-adiabatic field transition into the solenoid, results in the creation of a superposition quantum state, which generally has non-zero projections onto all 9 base functions. The superposition state evolves coherently in a reproduceable way which depends on $I_1$, this allows us to control the projection of the quantum state onto any of the $m_I, m_J$ basis states(*13*). The mass spectrometer illustrated in figure 1a measures the flux of $H_2$ molecules which undergo scattering into the specular channel for different total scattering angles ($\theta_{total}$) which can be set to either $45°$ or $22.5°$.

The red markers in figure 2a shows FD results for specular scattering of $H_2$ from a 500K Ni(111) surface. Scanning the magnetic field clearly modulates the scattered signal intensity, i.e. the quantum state of the molecule before it hits the surface changes the specular scattering probability. As the nuclear spin orientation is not expected to change the scattering probability, the oscillations we observe should reflect the scattering stereodynamics, i.e. the fact that different rotational projection states have different probabilities of scattering into the specular scattering channel. The oscillation interference patterns measured in previous MMI studies, which included state selection after scattering are sensitive to changes of both the magnitude and the phase of the molecular wave function during scattering, and their interpretation required fitting the elements of the scattering matrix(*14, 22, 27*). As we shall show below, the FD signal shown in figure 2a, which is insensitive to the final state of the molecule, allows us to employ a much simpler interpretation of the experimental results.

As the magnetic Hamiltonian for $J=1$, $I=1$ $H_2$ molecules is well known(*23*), we can use semi-classical calculations (See supplementary text) to propagate initially pure $m_I, m_J$ quantum states emerging from the end of the hexapole polariser and calculate the quantum state of the molecule just before colliding with the surface(*13*). Figure 2b shows the average populations of the beam particles as they arrive at the surface, using the surface normal as the quantisation axis. The populations in the $m_J=1,-1$ and 0 states as a function of the current $I_1$ are shown using red, blue and green lines respectively. Below we use the common terminology of calling molecules in the $m_J=1,-1$ states as the two (counter-rotating) helicopters, and those in the $m_J=0$ state as cartwheels.

One observation we can make by looking at the three populations plotted in figure 2b, is that the central ($I_1=0$) region is dominated by strong oscillations between the populations of the two counter-rotating helicopter molecules, which means that if these two states scattered with different probabilities we would expect our manipulation to produce strong oscillations at the centre, which is not what we see in the experimental data. This follows our expectations, as due to the symmetry of the surface with respect to the scattering plane, we anticipate the sense of helicopter rotation to not affect the scattering. Next we note that the experimental data in figure 2a has a similar pattern to that of the calculated cartwheel population (green line in figure 2b), suggesting cartwheel like molecules are more likely to scatter than those which approach the surface as helicopters.

To check this quantitatively we combined these two observations into a simple alignment scattering model in which scattering probabilities depend only on the alignment of the impinging molecules. We denote the equal scattering probabilities of both the two helicopter states ($m_J$=1,-1) as $P_H$ and that of the cartwheel ($m_J$=0) state as $P_C = \alpha P_H$ where $\alpha$ is the ratio of cartwheel to helicopter scattering probabilities. Within this model the FD signal can be written as
$$\frac{P_H N_1(I_1)+P_H N_{-1}(I_1)+P_C N_0(I_1)}{<P_H N_1(I_1)+P_H N_{-1}(I_1)+P_C N_0(I_1)>} = \frac{N_1(I_1)+N_{-1}(I_1)+\alpha N_0(I_1)}{<N_1(I_1)+N_{-1}(I_1)+\alpha N_0(I_1)>}$$ where $N_1$, $N_{-1}$ and $N_0$ are the populations in the $m_J$=1,-1 and 0 states as a function of the current, as shown in figure 2b. Note that we normalised the model to its average value along the entire $I_1$ range, similarly to what is done for the experimental data. The thick red line in figure 2a shows the model for the best fit enhancement factor $\alpha = 1.5$, which agrees remarkably well with the experimental data, especially considering there is only one free parameter in this comparison. It should be noted that due to the possibility of polarisation loses in our beam which lead to a background contribution to the signal, the value of $\alpha$ we quote should be treated as a lower limit, i.e. the scattering probability is at least a factor of 1.5 higher for cartwheel molecules in comparison to helicopter molecules.

The fact that the alignment scattering model follows the specularly scattered signal almost perfectly, and shows an enhanced scattering for cartwheel molecules, is a rather robust property of this system, as shown in figure 3a and 3b. Figure 3a compares FD measurements for a total scattering angle of 45° (red markers, identical to those shown in figure 2a) with those measured for a total scattering angle of 22.5° (blue markers). The pattern is similar, and again the comparison with the model (blue line) is very good with a slightly higher enhancement factor ($\alpha = 1.7$).

Figure 3b compares the results shown earlier in figure 2a (red asterisks) with results obtained with nozzle temperatures of 84K (black diamonds) and 149K (green triangles) changing the beam velocity to 1344ms$^{-1}$ and 1786ms$^{-1}$ respectively (supplementary text and figure S2). The change of the signal patterns follows the change in the population control pattern as evidenced by the fact that the alignment scattering model (blue and black lines in figure 3b) produce excellent fits to the data. For 84K we get a very similar enhancement factor ($\alpha = 1.4$) and for 149K we obtain a lower enhancement factor of $\alpha = 1.2$. We have also performed FD measurements for 3 different crystal azimuths in between $[10\bar{1}]$ and $[11\bar{2}]$, which all produce results which are identical within the experimental noise and are shown in figure S1a.

A completely different situation arises when we cool the surface (blue diamond markers in figure 3c), with the FD oscillations reducing by about an order of magnitude and reversing in polarity, i.e. helicopter molecules have a very small preference for scattering (best fit $\alpha = 0.97$, shown by the blue line in figure 3c). To understand the low surface temperature results we need to consider the dynamics of H$_2$ dissociation on this surface. When we expose the 500K nickel surface to our molecular beam, hydrogen molecules are constantly dissociating and sticking to the surface. However, at 500K the desorption rate is higher than the adsorption rate which means that the adsorbed H atoms quickly recombine and desorb from the surface, leaving an essentially clean surface for the molecules we measure, which are those that didn't dissociate and scattered elastically towards the detector. When Ni(111) is cooled below approx. 400K, the desorption and adsorption rates balance and H atoms remain on the surface with a coverage which depends on the flux of the molecular beam and on the surface temperature(28). When exposing a 180K surface to our beam, a complete saturation coverage is created and the surface becomes essentially inert.

The preferential cartwheel scattering observation, which we see when the surface is clean and reactive but disappears for the inert cold surface, could potentially be linked with stereodynamic trends of the dissociation reaction probability. Since diffractive scattering can, to a certain extent, be seen as the complementary channel to dissociation(*29*), enhanced scattering of cartwheel molecules could be linked with an enhanced reaction probability of helicopter molecules, as has been predicted from calculations of hydrogen reacting with other surfaces(*19*, *20*). However our observations can also be linked to opposing trends in other scattering channels which will be at the expense of the specular scattering probability.

In order to test whether there is a stereodynamic trend for reactive collisions, we need to combine our ability to control the rotational projection populations of the impinging molecules with a measurement that is sensitive to the coverage of H atoms on the surface. Low energy helium scattering is one of the most sensitive techniques for measuring particularly dilute adsorbate coverages(*30*). To perform the measurement, we seeded the $H_2$ beam with 10% helium ($^4$He). The majority gas ($H_2$) reacts with the surface and can be stereodynamically controlled by changing the current in the solenoid, whereas the helium atoms in the beam are used to monitor the surface coverage through its effect on the reflectivity of the surface. Figure 1b illustrates this mode of measurement where two values of $I_1$ were chosen to maximise the difference between the calculated helicopter/cartwheel populations. Since the helium atoms ($^4$He) have a zero nuclear spin and are completely unaffected by magnetic fields, any change in their scattered intensity can only reflect a change of the surface itself, i.e. the number of H atoms present on the surface.

To be able to follow changes in the surface coverage we set the surface temperature to 375K. As discussed in more detail in the supplementary text, at this temperature equilibrium is achieved between adsorption from the $H_2$ beam and thermal desorption for a surface coverage of approximately 0.06 monolayers (ML), a coverage which reduces the helium reflectivity by 37.5±1.5% in comparison to the clean Ni(111) surface. Figure 4a shows the scattered helium signal within a 170s window, where the control solenoid current was set to abruptly change between $I_1$=-0.018A and $I_1$=-0.0207A as shown in figure 4b, values which were chosen to minimise and maximise the helicopter populations. We note that adding 10% helium to the beam slowed the $H_2$ molecules by approximately 100ms$^{-1}$, which shifts the two population control currents by 0.0013A (supplementary text and figures S3 and S4) with respect to the pure $H_2$ beam. The measurements shown in figure 4a were repeated 150 times to obtain sufficient signal to noise and to calculate the standard deviation of the measured values. The helium signal, which was normalised to its average value, follows the magnetic manipulation sequence, decreasing when we enrich the beam with helicopters (68%) and increasing when we reduce the helicopter population (59%). Figure 4c shows an average of the measurement points at each of the two currents, excluding the first 8 seconds after changing the current $I_1$ to allow the surface coverage to stabilise. The helium reflectivity changes by ~1% between the two $I_1$ values.

The helium reflectivity results shown in figure 4, show that our magnetic manipulation controls not only the $H_2$ scattering channel but also the reaction channel, i.e. enriching the $H_2$ beam with more helicopter states (68% instead of 59%) leads to a larger yield of reaction products, i.e. more H atoms adsorbed on the surface. To quantify the difference in sticking probability we relate the changes in the helium reflectivity to the surface coverage of H atoms (supplementary text) and obtain an estimation of 1.2 for the minimum ratio between the sticking coefficients of helicopter and cartwheel molecules.

One possible explanation for the anticorrelated stereodynamic trends we measured, i.e. the increased scattering probability of cartwheel molecules and the increased reactivity of helicopter molecules is that the scattering event, which takes place further away from the surface and correspondingly before the molecule is close enough to react, acts as a stereodynamic filter to the reaction, reflecting more cartwheel molecules back into the gas phase and reducing their contribution to the dissociative adsorption reaction channel. To confirm this hypothesis would require further measurements of the stereodynamic trends of other possible competing channels (e.g. scattering into other diffraction peaks, diffuse scattering and bound state resonances). Alternatively, insight into the mechanism could be obtained by analysing trajectories from calculations which can reproduce the trends we measured. It is also interesting to note that a steering mechanism, which seems to be important for collisions of low energy $H_2$ molecules with other surfaces(*31–33*), would be expected if strong enough to eliminate the dependence on the rotational orientation of the impinging molecules, in contradiction to what we observe experimentally. Finally, the fact that the scattering stereodynamic trends do not change between 375K and 600K (figure S1b) means that the quantitative observations of this study, could be used to benchmark state-of-the-art calculations which rely on the static surface approximation.

**Acknowledgments:** The authors thank Geert-Jan Kroes and Mark Somers for valuable advice and acknowledge the support of the Supercomputing Wales project, which is part-funded by the European Regional Development Fund (ERDF) via Welsh Government.

**Funding:**

UKRI, Future Leader Fellowship MR/X03609X/1 (H.C)

EPSRC, grant EP/X037886/1 (GA, HC).

**Competing interests:** Authors declare that they have no competing interests.

**Data and materials availability:** All data are available in the main text or the supplementary materials.

**Supplementary Materials**

Supplementary Text

Figs. S1 to S6

References (*34–39*)

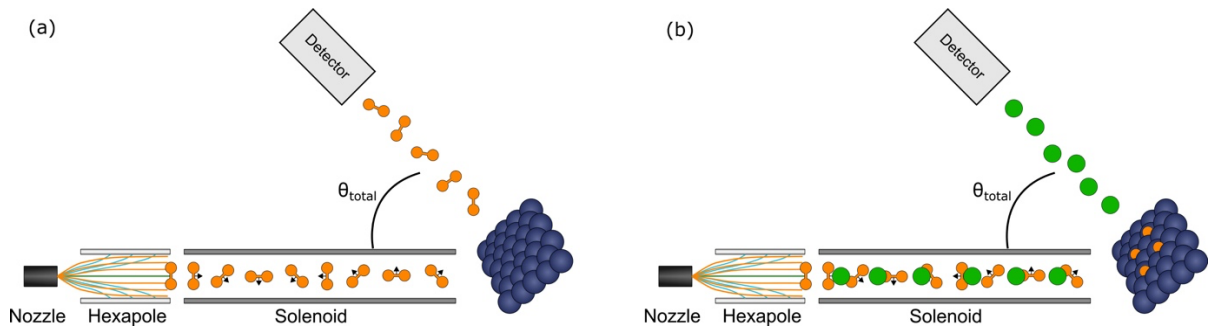

**Fig. 1. Illustration of the measurement method. (a)** Schematic of the setup used for the FD measurements. The rotational orientation of the molecules which reach the surface is controlled by a combination of an initial hexapole followed by a perpendicular field $B_1$, generated by passing the control current $I_1$ through a solenoid coil. The controlled beam then collides with a surface. The flux of $H_2$ molecules scattered in the specular direction is measured by mass spectrometers, positioned at $\theta_{total} = 45°$ or $22.5°$. **(b)** To measure sticking stereodynamics a small fraction (10%) of helium (illustrated as green circles) is mixed in the $H_2$ beam. By switching between two currents in the solenoid ($I_1$), the relative populations of helicopter and cartwheel molecules reaching the surface are modulated. The flux of helium atoms scattered from the surface into a mass spectrometer is used to monitor the surface coverage and how it changes for different $I_1$ values.

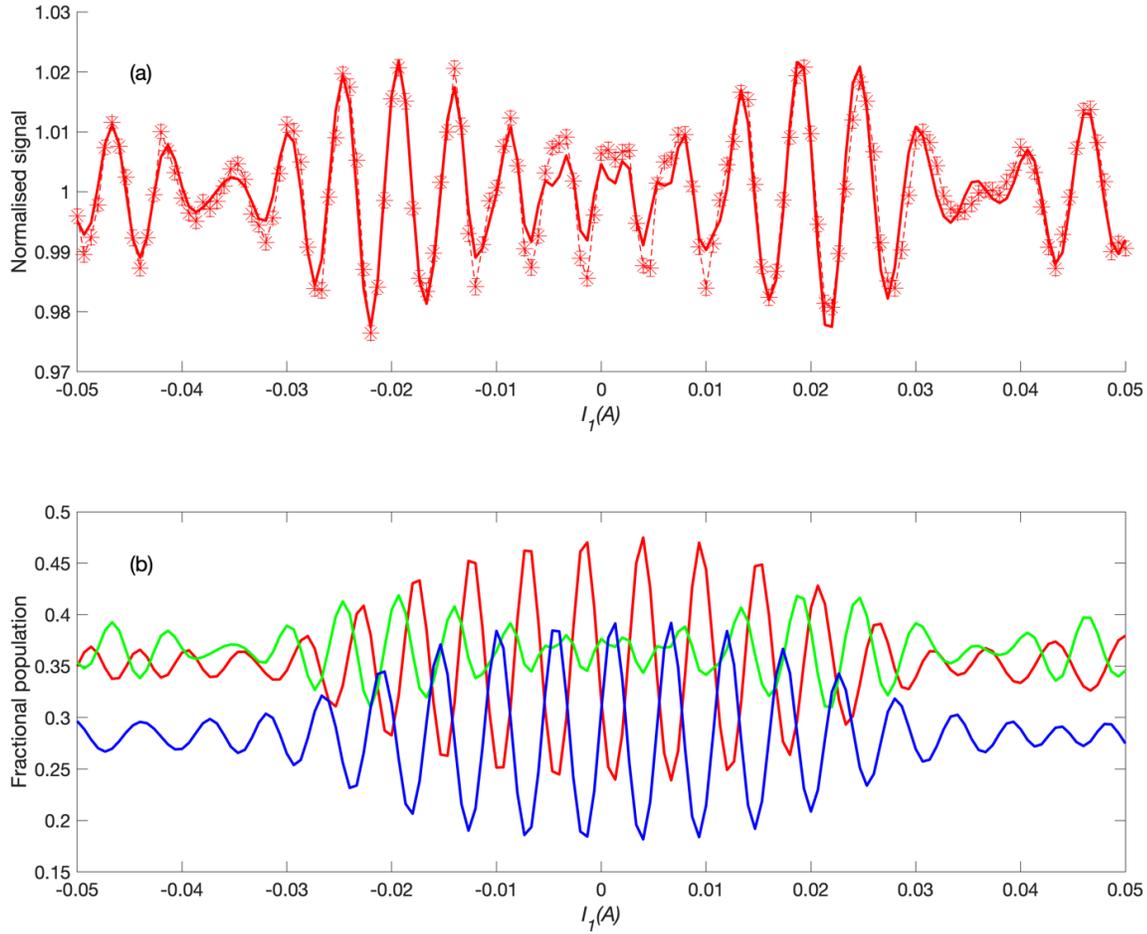

**Fig. 2. Stereodynamic control of the specular scattering flux. (a)** The red asterisk markers connected by red dashed lines show the normalised signal of a flux detection measurement from a 500K Ni(111) surface as a function of the solenoid current, $I_1$. The error bars show the standard deviation of the values calculated from repeat measurements. The thick red line shows the alignment scattering model for $\alpha = 1.5$, which fits the data almost perfectly. **(b)** The red, blue and green lines show the calculated average populations in the $m_J=1,-1$ and 0 states of the beam arriving at the sample as a function of the control current $I_1$. The surface normal was used as the quantisation axis

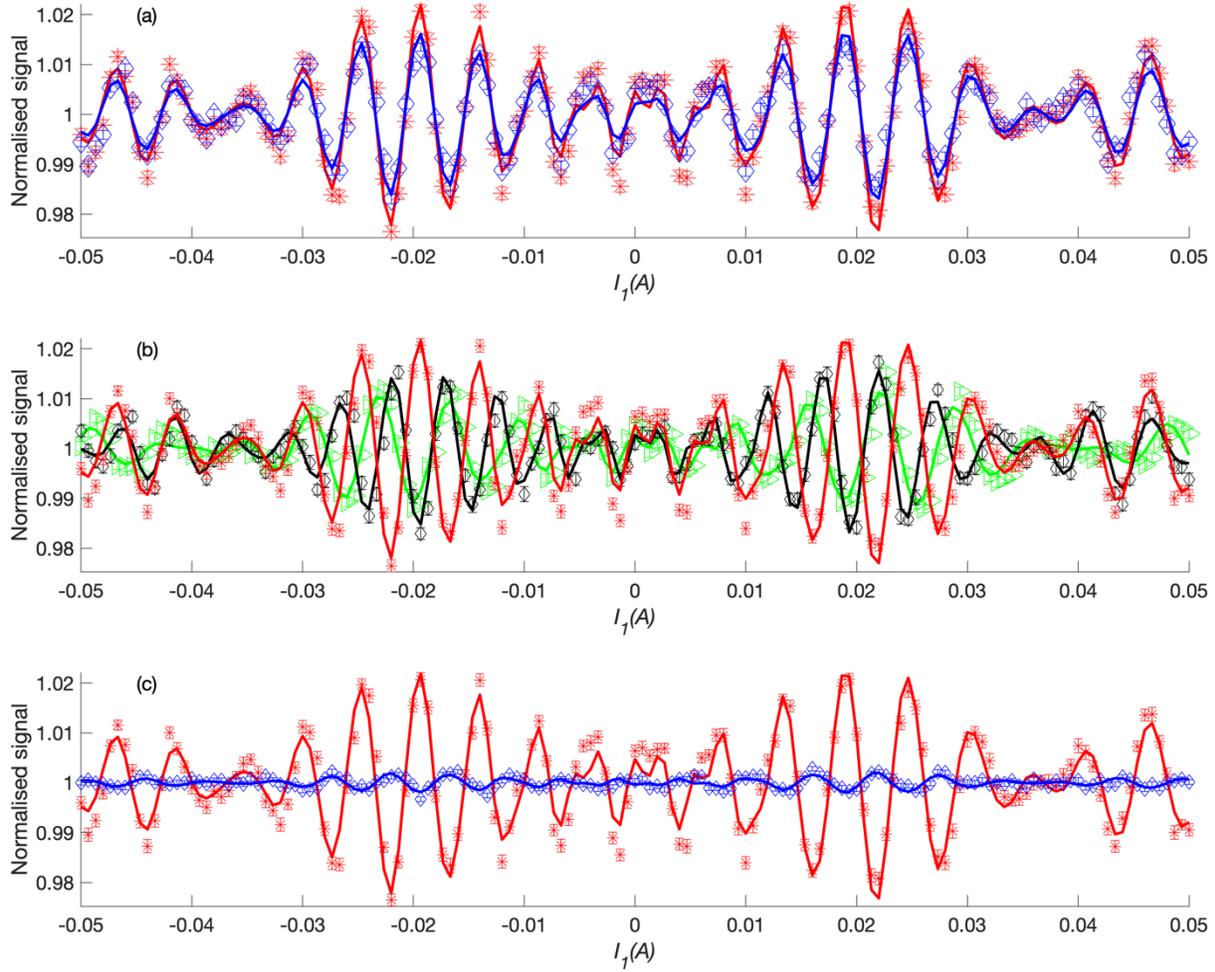

**Fig. 3. The dependence of the specular scattering flux on the geometry, beam velocity and surface temperature.** *(a) Comparing FD scattering measurements from a 500K surface, using a 106K nozzle. Red and blue markers and are for two different scattering geometries $\theta_{total} = 45°$ and $22.5°$. The alignment scattering model fits are shown as lines with corresponding colours. (b) Comparing FD scattering measurements from a 500K surface for $\theta_{total} = 45°$, using a 106K nozzle (red), an 84K nozzle (black) and a 149K nozzle (green). The fits to the alignment model are shown as lines with corresponding colours. (c) Comparing FD scattering measurements using a 106K nozzle and $\theta_{total} = 45°$ for a 500K surface (red) which is clean and reactive to FD measured from a 180K surface covered with a passivating layer (blue). The alignment scattering model fits are shown as lines with corresponding colours.*

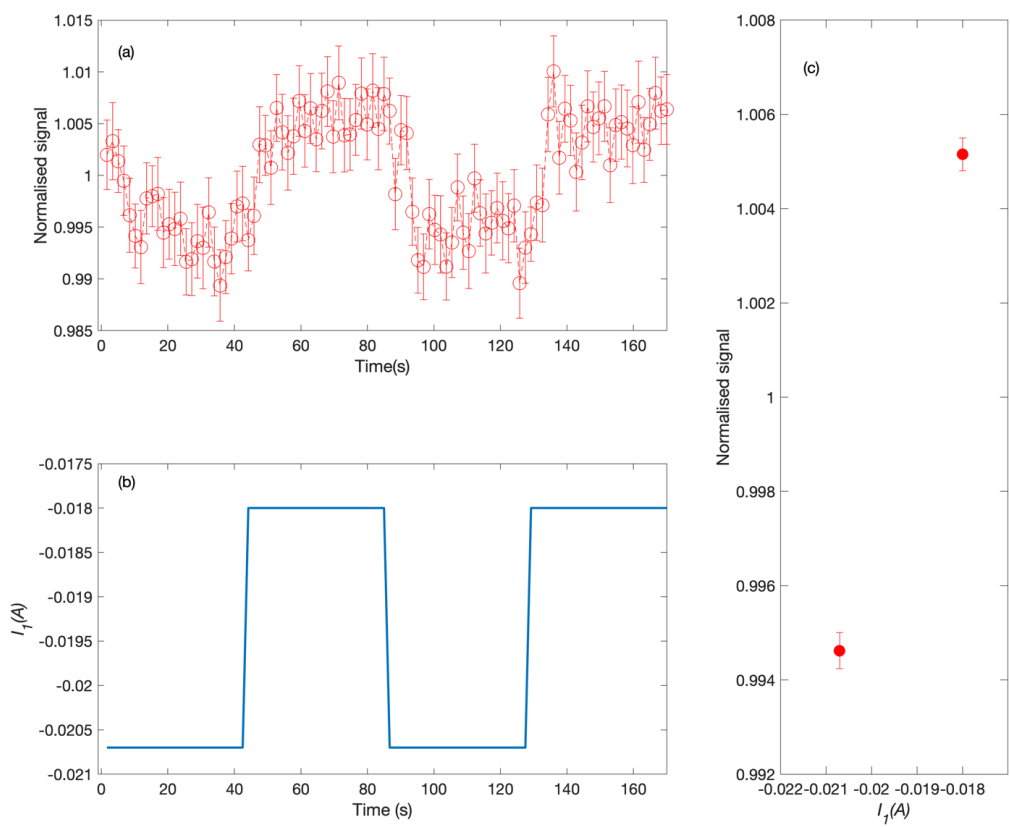

**Fig. 4. Stereodynamic control of $H_2$ sticking, monitored by helium scattering.** *(a) Normalised helium signal as a function of time. (b) Solenoid control current, $I_1$, as a function of time. (c) Average helium signals for the two control currents, excluding the first 8 seconds after changing the current to let the surface coverage stabilise. The error bars represent ±1 standard deviation.*

**Supplementary Materials for: Anticorrelated stereodynamics of scattering and sticking of H₂ molecules colliding with a reactive surface**

H. Chadwick, G. Zhang, C.J. Baker, P.L.Smith and G. Alexandrowicz

Corresponding author: g.n.alexandrowicz@swansea.ac.uk

**The PDF file includes:**

Materials and Methods

Supplementary Text

Figs. S1 to S6

**Materials and Methods**

**Experimental Methods**

A modified version of the magnetic molecular interferometer (MMI) apparatus(*14*) was used in the current work as outlined below. Below we briefly reiterate the basic principles of the setup and provide details for the elements which are specific to this study. The molecular beam expansion was created using a nozzle with the coldest central part of the beam selected using a skimmer. Whilst both the $J = 1, I = 1$ and $J = 0, I = 0$ states are populated, the para-H₂ ($J = 0, I = 0$) cannot be magnetically manipulated and adds a small constant background to the measured signals.

The molecules then enter a magnetic hexapole(*26*), which creates an inhomogeneous magnetic field where the different $m_I, m_J$ states are either focussed or defocussed depending on their magnetic moment(*23*). At the end of the hexapole, there is a hexapole to dipole transition which adiabatically rotates the states to a common quantisation axis defined by the direction of the dipole.

After the dipole, the H₂ molecules travel down the remainder of the first arm of the beamline, first passing through a field free region before entering the solenoid which creates a magnetic field in the opposite direction to that which the molecules are travelling in. The solenoid currents are controlled within a ppm accuracy using a Danfysik 0 – 10A power supply. After the solenoid the $H_2$ continues through the remainder of the first arm of the beamline, which also contains a small, current independent, magnetic field, before colliding with the surface.

The Ni(111) sample (surface preparation lab) was mounted on a home-built, 6-axis manipulator in the ultra-high vacuum chamber at the end of the first arm of the apparatus. The surface was cleaned using repeated sputter-anneal cycles, where sputtering was performed by exposing a 500K surface to a ~12µA beam of 1 KeV argon ions for 10 minutes, annealing was performed by heating the sample to 800K for 15-30 seconds, and the cycles were repeated until the specular scattering signal stopped improving. The temperature of the sample was monitored using a T-type thermocouple with an absolute error estimated as $\pm 0.75$K. The surface azimuth orientation was verified to within $\pm 1°$ using the helium diffraction pattern.

Two differentially pumped mass spectrometers at different scattering angles were used to detect the scattered $H_2$ beam. The first is an SRS RGA 200 connected to a port on the UHV chamber, at a total scattering angle of 22.5° (corresponding to an incoming and outgoing angle of 11.25° for the specularly scattered measurements presented here). The second is a Hiden HAL 201 which can be moved into the beamline after the solenoid in the second arm of the MMI apparatus, which detects signal scattered through a total scattering angle of 45° (22.5° incoming and outgoing angle for specular scattering). To monitor the scattered helium signal for the reactivity measurement, an Extrel Max120 was used which was also positioned at a total scattering angle of 45° but offered higher sensitivity.

Theoretical methods

To obtain the $m_J$ state populations of the molecules that collide with the sample, it is necessary to calculate the propagation of the 9 $m_I$, $m_J$ states of $H_2$ through the magnetic components of the beamline shown schematically in figure 1A of the main manuscript. The first step is to calculate the probabilities, $P_{hex}(m_I, m_J)$, that each $m_I$, $m_J$ is transmitted through the hexapole polariser. To do this we use a semi-classical trajectory calculation where the motion of the molecule is propagated classically, but the (quantised) forces are calculated according to their $m_I$, $m_J$ state(34). Due to the strong magnetic field gradients in the hexapole, the superposition states decohere(35) leaving the molecules in 1 of the 9 pure $m_I$, $m_J$ states at the end of the hexapole with an unequal population distribution.

The propagation of the quantum states from the end of the hexapole to the surface is also performed semi-classically(13), with the motion of the centre of mass of the molecule calculated classically but the evolution of the $m_I$, $m_J$ states quantum mechanically, using the Hamiltonian for the $J = 1$, $I = 1$ state of $H_2$(23). The solution of the evolution can be written as a propagation matrix, $U(I_1)$, which expresses the superposition state obtained from an initially pure $m_I$, $m_J$ state after it was propagated through the magnetic field profile associated with a given solenoid current, $I_1$. The relative population in a given $m_J$ state at the surface ($\Omega(m_J')$) can then be calculated by projecting the wavefunction of the molecule on to the $m_I'$, $m_J'$ state at the surface (denoted by ' and where the quantisation axis is taken to be the surface normal) and taking the square modulus before summing over the final $m_I$ states, velocity distribution ($P_v$) and hexapole transmission probabilities, i.e.,

$$\Omega(m_J')=\sum_{v} P_v \sum_{m_I'} P_{hex}(m_I,m_J)|\langle m_I',m_J'|R(\theta)U(I_1)|m_I,m_J\rangle|^2$$

where $R(\theta)$ is the rotation matrix that changes the quantisation axis from the direction of the dipole at the end of the hexapole to the surface normal.

Supplementary Text

Determination of the molecular beam velocities and control currents

For the $T_N$ = 84K, 106K and 149K data measured with a pure $H_2$ beam (figures 2 and 3), the velocities were obtained from full interferometer measurements where a second hexapole positioned before the detector is used to perform state-to-state scattering measurements(*14*). The value of the current in the solenoid of the second arm of the apparatus was set to $I_2$ = 0A. The measured oscillation curves, presented in the top panel of figure S2, were fit using a procedure described previously(*22*) where parameters including the central velocity and full width at half maximum (FWHM) of the velocity distribution (which is modelled as a gaussian) are allowed to vary. A comparison of the velocity distributions that were obtained are presented in the bottom panel of figure S2. For the $T_N$ = 84K measurement, the central velocity is 1344 ms$^{-1}$ with a FWHM of 6.9%, the $T_N$ = 106K velocity is 1513ms$^{-1}$ with a FWHM of 7%, and at $T_N$ = 149K the velocity is 1786 ms$^{-1}$ with a FWHM of 7.8%.

A slightly different procedure was used for the 10% He in $H_2$ reactivity measurement, where seeding the beam with helium leads to a small reduction of the mean $H_2$ velocity. In this case, the oscillations in the short FD measurement that was made to determine the optimal control currents to use in the reactivity measurement were scaled to find the velocity that best matched the oscillation frequency of the signal. The top panel of figure S3 presents the flux detection measurement measured for a pure $H_2$ beam (black) and the 10% He in $H_2$ mix (red), and the bottom panel the resulting velocity distributions. This gave a central velocity of 1410ms$^{-1}$ and FWHM of 8% for the 10% He seeded $H_2$ beam.

The slight reduction in the $H_2$ velocity when the beam is seeded with He meant that the values of the magnetic field which maximise the 'helicopter' and 'cartwheel' populations in the two beams need to be shifted by 0.0013A with respect to the optimal values for the pure beam. Calculations were performed analogously to those shown in the bottom panel of figure 2 of the main manuscript for the velocity distribution obtained for the seeded beam, to determine how the $m_J$ state populations changed as a function of magnetic field for this different velocity distribution. The results of these calculations are presented in figure S4.

Estimation of the coverage during the sticking measurements

To determine the equilibrium coverage during the reactivity measurements presented in figure 4, we monitored the transient change in the specularly scattered helium intensity when a clean surface is exposed to the beam. We used the same conditions for the molecular beam which were used in the rotationally controlled sticking measurements (Tn=106K, 10% He in $H_2$ mixture and a surface temperature of 375K). A separation valve along the beam line was opened abruptly to allow the molecular beam to hit the surface and the drop in helium signal due to the adsorption of H atoms was recorded, until it had plateaued, after which the separation valve was then closed. The results of this measurement are shown in figure S5.

Exposing the surface to the mixed beam leads to a decay of the helium signal to 62.5±1.5% of its value before H atoms were adsorbed on the surface. To relate this decay to the H atom coverage, we performed a second set of experiments at a low enough temperature where desorption is negligible (220K) and we can follow the growth of the surface layer up to the formation of an ordered structure at a coverage of 0.5ML(*36*). The layer was grown by back-filling the UHV chamber with a $H_2$ pressure of $5 \times 10^{-7}$mbar and monitoring how the scattering intensity of a molecular beam of He changes as a function of hydrogen dose. This is shown in the top panel of figure S6 for two repeat measurements where scattering into the specular channel was monitored (black and red), and one where the scattering into the ½ order diffraction channel was measured. To convert from hydrogen dose (ε) to hydrogen coverage (θ), it was assumed that the sticking coefficient, S(θ), decreased linearly from 0.04 to 0 for coverages between 0 and 0.5 monolayer. i.e., $S(\theta) = \frac{d\theta}{d\varepsilon} = 0.08(0.5 - \theta)$. Previous studies have shown that the initial sticking coefficient is on this order(*16, 37–39*) for the beam energies we used in this study. The differential equation for $\frac{d\theta}{d\varepsilon}$ gives us the coverage as a function of dose ($\theta = 0.5(1 - \exp(-0.08\varepsilon))$), which can then be used to convert the $H_2$ dose in the back-filling measurement to coverage, from which the dependence of the helium intensity on hydrogen coverage immediately follows. The result of this is presented in the bottom panel of figure S6 for each of the intensity measurements presented in the top panel, where the intensity of the measured signals, $\Lambda$, have been normalised to the maximum intensity, $\Lambda_0$. This conversion produces a maximum in the ½ order diffraction peak scattered intensity at a coverage of approximately 0.5 monolayers, in agreement with that found in previous work(*36*).

Assuming that the drop in the relative signal intensity due to adding H atoms on the surface is independent of the surface temperature, we can use the 37.5% intensity drop from the first measurement to estimate the hydrogen coverage on the surface as 12±0.5% of a monolayer, as shown by the grey line in the bottom panel of figure S6. Whilst this is a relatively crude estimation, it is in reasonable agreement with previously obtained values(*37*) at similar surface temperatures and $H_2$ pressures.

Estimation of the ratio of the 'helicopter' and 'cartwheel' sticking coefficent

The relatively low equilibrium coverage the sticking measurement was performed at allows us to use a simple linear model to relate the attenuation of the helium signal to the adsorbate coverage(*30*), $\frac{\Lambda}{\Lambda_0} = 1 - \beta\theta$, where $\beta$ is a constant (see bottom panel of figure S6). The coverages from the reactivity measurement can then be calculated from the values of $\frac{\Lambda}{\Lambda_0}$ that were obtained as $\theta = (1 - \frac{\Lambda}{\Lambda_0})/\beta$. Taking the ratio of the two coverages at the two solenoid current values which we will denote $I_{1a}$ and $I_{1b}$, and normalising the signal such that $\Lambda_0 = 1$ gives $\frac{\theta(I_{1a})}{\theta(I_{1b})} = \frac{1-\Lambda(I_{1a})}{1-\Lambda(I_{1b})}$. The values of $\Lambda(I_{1a})$ and $\Lambda(I_{1b})$ can be found from applying the ±0.5% modulation seen in figure 4c to the attenuation of the helium signal when we do not modulate the populations ($\frac{\Lambda}{\Lambda_0} = 0.625$, shown in figure S5) resulting in $\frac{\theta(I_{1a})}{\theta(I_{1b})} = 0.98$.

On the other hand the coverage at the two control currents can also be calculated as the product of the the flux of the beam and the sticking coefficients of the different states, i.e., $\theta(I_1) = F[S_H N_{-1}(I_1) + S_H N_1(I_1) + S_c N_0(I_1)]$ where $N_{m_J}(I_1)$ is the proportion of the beam in a given $m_J$ state and at a given solenoid current ($I_1$), $S_H$ is the sticking probability for helicopter ($m_J = \pm 1$) molecules, $S_c$ is the sticking probability for cartwheel ($m_J = 0$) molecules and F is the flux of the molecular beam. Defining a ratio for the sticking coefficients of helicopter and cartwheel molecules, $\alpha' = \frac{S_H}{S_c}$, we can equate the two expressions for the coverage ratio $\frac{\alpha'[N_{-1}(I_{1a})+N_1(I_{1a})]+N_0(I_{1a})}{\alpha'[N_{-1}(I_{1b})+N_1(I_{1b})]+N_0(I_{1b})} = \frac{\theta(I_{1a})}{\theta(I_{1b})} = 0.98$. Using the calculated populations at the two currents (figure S4) we obtain that the ratio for the sticking coefficients of helicopter to cartwheel molecules is $\alpha' = 1.2$.

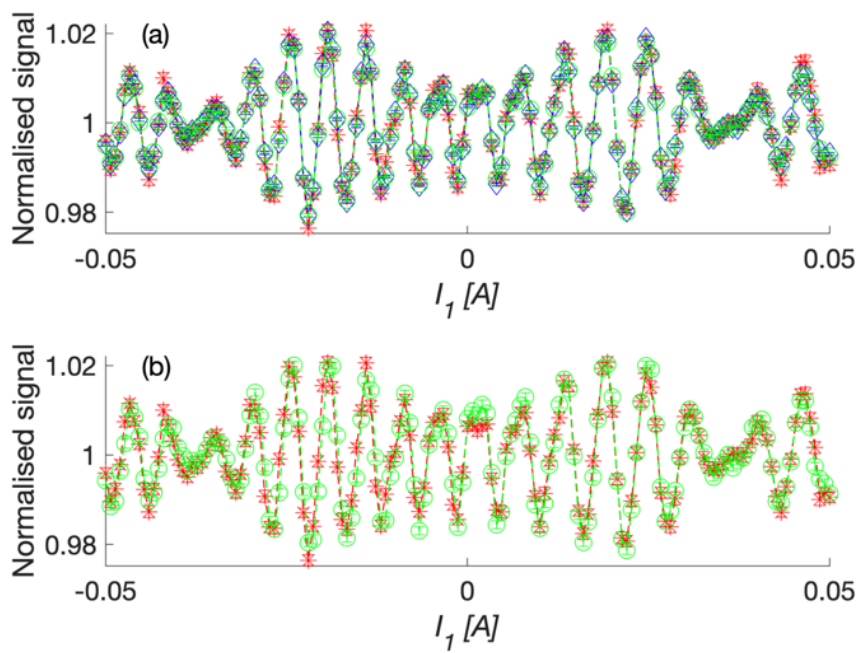

**Fig S1**. (a) Comparison of flux detection measurements performed at different crystal azimuths for a 500K surface temperature and a 106K nozzle temperature. The red asterisk and green circle markers show the results measured along the $[11\bar{2}]$ and $[10\bar{1}]$ directions respectively, whereas the results when measuring in between these two azimuths is plotted using the blue diamond markers. All of the results are identical within the experimental uncertainties. (b) Comparison of flux detection measurements performed at two different temperatures where the surface is still reactive, The red asterisk and green circle markers show the results for surface temperatures of 500K and 375K respectively, using the same nozzle temperature (106K) and measuring along the same crystal azimuth ( $[11\bar{2}]$). The results are identical within the experimental uncertainties.

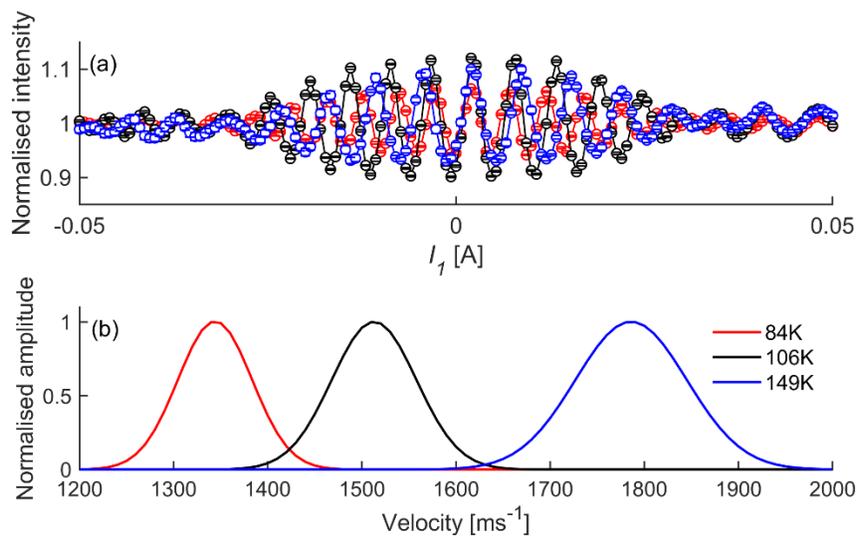

**Fig S2.** (a) Comparison of full-interferometer oscillation curve measurements performed at different nozzle temperatures. The red markers correspond to a nozzle temperature of 84K, the black to 106K and the blue to 149K. The surface temperature was 500K, and the second solenoid current 0A. (b) Comparison of the velocity distributions for the different nozzle temperatures. The red line corresponds to the velocity distribution obtained by fitting the red oscillation curve in panel (a) measured at a nozzle temperature of 84K, the black to 106K and the blue to 149K.

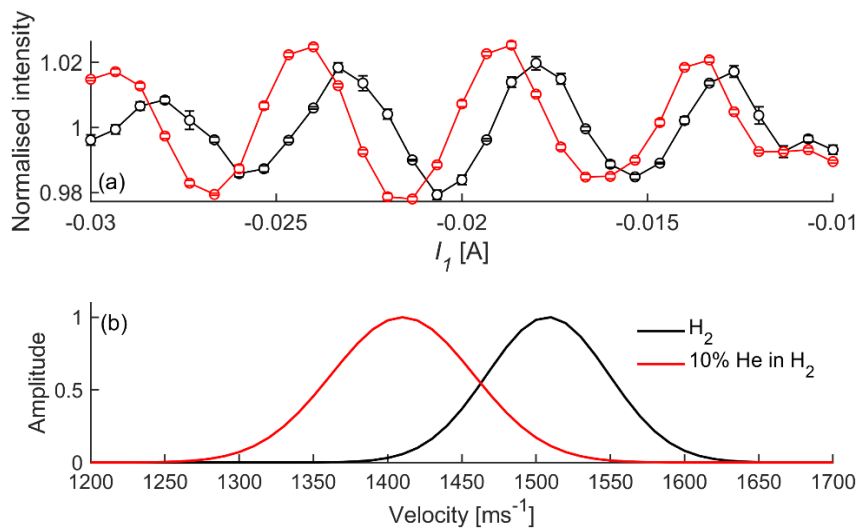

**Fig. S3**. (a) Comparison of flux-detection oscillation curves measured for two different incident molecular beam conditions. The black markers correspond to a pure $H_2$ molecular beam, and the red to the 10% He in $H_2$ molecular beam. (b) Comparison of the molecular beam velocity distributions obtained for the pure $H_2$ molecular beam and the 10% He in $H_2$ molecular beam. The same colours are used as for panel (a).

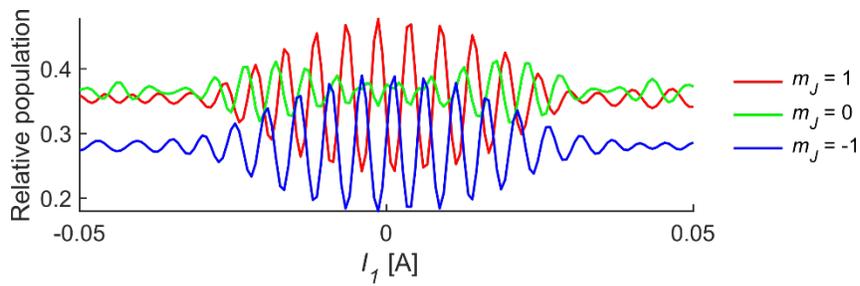

**Fig. S4.** Calculated $m_J = 1$ (red), $m_J = 0$ (green) and $m_J = -1$ (blue) state populations at the surface as a function of solenoid current, using the reduced velocity for $H_2$ when 10% He is added to the beam. The surface normal was used as the quantisation axis.

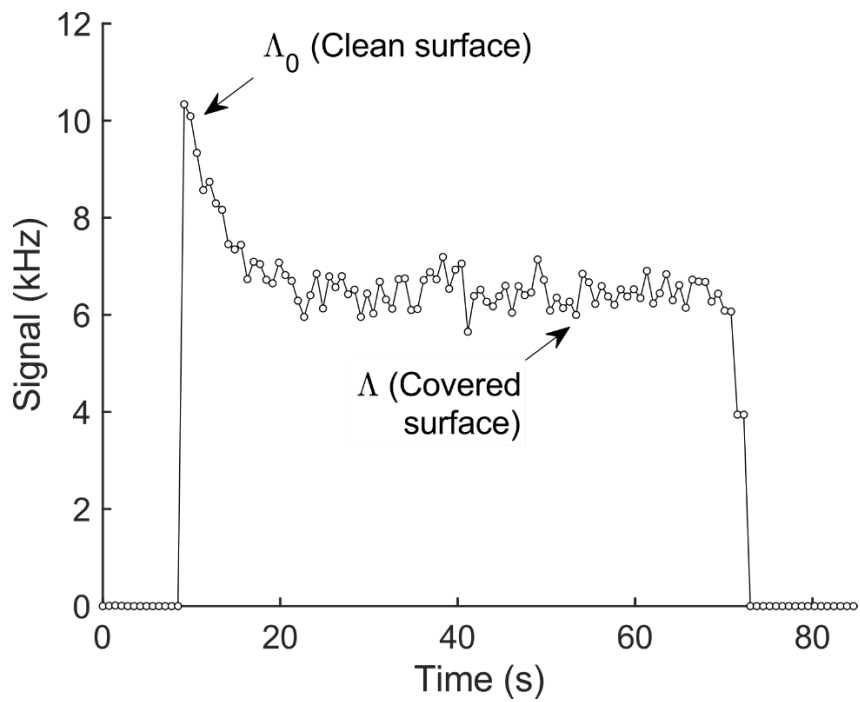

**Fig. S5.** Specularly scattered $^4$He signal obtained when opening and closing a separation valve for a 10% He in $H_2$ mix colliding with a Ni(111) surface held at a surface temperature of 375K.

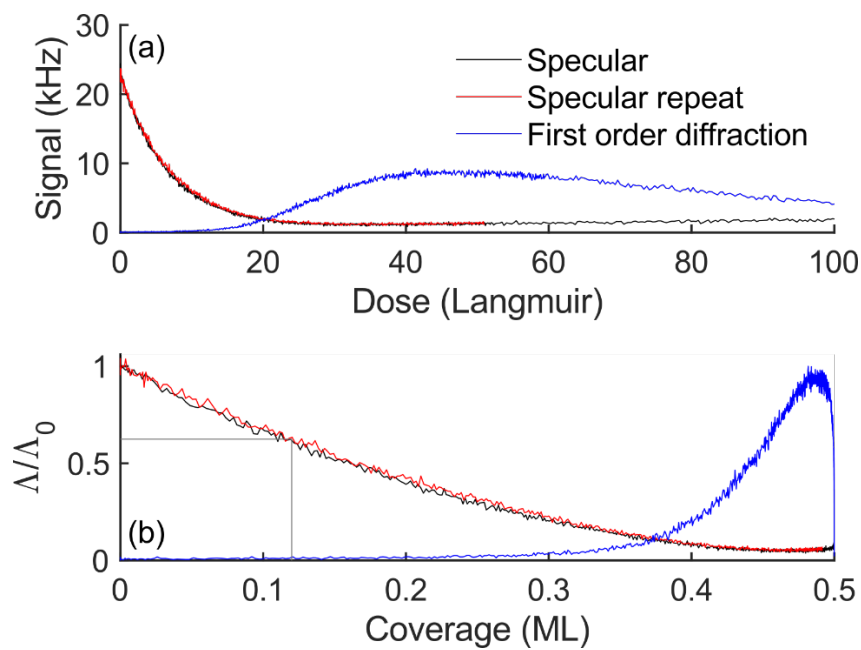

**Fig S6. (a)** Comparison of scattered helium intensity as a function of hydrogen dose during various uptake measurements performed by backfilling the UHV chamber with $H_2$ at a pressure of $5\times10^{-7}$ mbar at a surface temperature of 220K. The black and red lines both show a measurement monitoring the $^4$He scattering intensity on specular, and the blue line the $^4$He intensity scattered into a first order diffraction peak. (b) Normalised $(\Lambda/\Lambda_0)$ $^4$He scattered signal as a function of H coverage ($\theta$) on the surface obtained from the measurements shown in panel (a). The two specular measurements are again shown in black and red, and the diffraction peak in blue. See text for details.